# Equity Markets Volatility, Regime Dependence and Economic Uncertainty: The Case of Pacific Basin

Bahram Adrangi
W.E. Nelson Professor of Financial Economics
University of Portland
5000 N. Willamette Blvd.
Portland, Oregon 97203
adrangi@up.edu

Arjun Chatrath
Schulte Professor of Finance
University of Portland
5000 N. Willamette Blvd.
Portland, Oregon 97203
chatrath@up.edu

Saman Hatamerad
University of Zanjon
samanhatamerad@yahoo.com

Kambiz Raffiee
Foundation Professor of Economics
College of Business University of Nevada, Reno
Reno, Nevada 89557
raffiee@unr.edu

March 2025

# Equity Markets Volatility, Regime Dependence and Economic Uncertainty: The Case of Pacific Basin

Abstract

This study investigates the relationship between the market volatility of the iShares Asia 50 ETF (AIA) and economic and market sentiment indicators from the United States, China, and globally during periods of economic uncertainty. Specifically, it examines the association between AIA volatility and key indicators such as the US Economic Uncertainty Index (ECU), the US Economic Policy Uncertainty Index (EPU), China's Economic Policy Uncertainty Index (EPUCH), the Global Economic Policy Uncertainty Index (GEPU), and the Chicago Board Options Exchange's Volatility Index (VIX), spanning the years 2007 to 2023. Employing methodologies such as the two-covariate GARCH-MIDAS model, regime-switching Markov Chain (MSR), and quantile regressions (QR), the study explores the regime-dependent dynamics between AIA volatility and economic/market sentiment, taking into account investors' sensitivity to market uncertainties across different regimes. The findings reveal that the relationship between realized volatility and sentiment varies significantly between high- and low-volatility regimes, reflecting differences in investors' responses to market uncertainties under these conditions. Additionally, a weak association is observed between short-term volatility and economic/market sentiment indicators, suggesting that these indicators may have limited predictive power, especially during high-volatility regimes. The QR results further demonstrate the robustness of MSR estimates across most quantiles. Overall, the study provides valuable insights into the complex interplay between market volatility and economic/market sentiment, offering practical implications for investors and policymakers.

**Equity Markets Volatility, Regime Dependence and Economic Uncertainty: The Case of Pacific Basin**

**1. Introduction**

Behavioral finance literature (e.g., Baker and Wurgler (2006, 2007)) holds that investors are often irrationally driven by excessive unjustifiable pessimism or optimism, rather than fundamental factors. Thus, asset price dynamics may also be explained by investor psychology, contrary to thesis or efficient markets (Fama (1965)). This paper addresses issues regarding the role of investor sentiment or uncertainty in short-term volatility or the persistence of long-term volatility by examining the popular Ishares Asia 50 ETF, AIA. AIA component companies are active in all sectors of several Asian economies with roughly 35 percent in technology, 25 percent in financial, and 17 percent in consumer cyclical sectors, respectively.

Understanding equity market volatility is crucial in various financial applications, such as portfolio optimization, risk assessment (e.g., stress testing), and derivatives pricing. In the context of futures contracts, volatility serves as a key factor contributing to phenomena like contango and backwardation, as explored by researchers such as Robe and Wallen (2016), Beckmann and Czudaj (2014), and Alizadeh and Nomikos (2011). Others like Koulis and Kyriakopoulos (2023), among others focus on volatility spillovers among asset classes and commodities, as well as precious metals like gold and silver. .

There are difficulties in precisely measuring the volatility of the equities market, nevertheless, as several studies have shown. According to Peters (1994), Eve et al. (1997), and Adrangi et al. (2001), equity price dynamics frequently follow nonlinear processes, show clustering, and fluctuate over time (e.g., Bollerslev et al. (1988)). Additionally, research by Ang and Chen (2002) and Ang and Bekaert (2015) has shown that co-movements across asset classes typically get more intense during times of market turmoil.

By investigating the possible relevance of investor mood in nonlinear volatility dynamics, this work adds to the body of previous literature. In particular, we look at how factors that reflect opinions and uncertainties about US markets, China's economic policy, and international economic policy connect to the volatility of the Asian equities market (represented by AIA) between 2007 and 2023.

We use Baker et al. (2012a, 2012b, 2014, 2016, , 2019, 2021) indices to measure financial and economic risks. These indices offer important insights into the complex processes of equity market volatility by capturing uncertainties in both US and international markets.

The manner in which businesses are impacted by economic instability was not explored theoretically or empirically a few decades ago. Early pioneers like Sandmo (1971) sowed the roots for this investigation, which were later emphasized by seminal publications like Galbraith's "The Age of Uncertainty" (1977).

Scholars such as Bernanke (1983), Pindyck (1991), Dixit & Pindyck (1994), Flacco & Kroetch (1986), Fooladi & Kayhani (1990, 1991), Golchin and Riahi (2021), and Adrangi & Raffiee (1999) have built upon this foundation by investigating how businesses adjust to uncertainty in areas such as production, pricing, investments, and profit-seeking. They sought to comprehend how companies manage market risks and make choices during ambiguous periods.

The impact of "uncertainty shocks" in causing economic cycle fluctuations is highlighted by more recent research, including studies by Bloom (2007, 2009, 2014), Fernández-Villaverde et al. (2011), Basu & Bundick (2017), Jurado et al. (2015), and Baker et al. (2016). Interestingly, Baker et al. (2016) employed text analysis to show that policy uncertainty serves as a warning indicator for falling output, employment, and investment in major economies, including the US.

Through index construction, Baker et al.'s (2012a, 2014, 2016) work has been crucial in assessing and analyzing economic and economic policy uncertainty during the last 20 years. Their work helps investors, economists, and policymakers better grasp these crucial elements by providing insightful information about the dynamics of economic uncertainty.

According to Baker et al. (2016), governmental policy uncertainty peaked in the wake of the 2008 global financial crisis. The recovery from the crisis was severely hampered by this uncertainty, which resulted from worries about the government's future position on spending, taxation, monetary policy, healthcare, and regulatory frameworks. This increased policy uncertainty caused businesses and people to postpone decisions about investments and consumption expenditures.

Many indicators have been developed as a result of researchers' heavy focus on assessing different facets of mood and uncertainty. While Baker et al. (2012a, 2012b, 2014, 2016, 2019, 2021) developed a number of indices, including Twitter-based measures of economic uncertainty (TEU), Manela and Moreira (2017) presented the NVIX, a monthly news-driven indicator. According to research by Su et al. (2017, 2018, 2019), Altig et al. (2020), Krol (2014), Dutta et al. (2021), Jiang et al. (2019), Pan et al. (2021), and Lindblad (2017, 2019), Adrangi & Hamilton (2023), Adrangi et al. (2015), Adrangi et al. (2024a, 2024b), Bonaime et al. (2018), Im et al. (2017), among others. these indicators are now essential for assessing volatility across various asset classes.

Together, the results of these research provide credence to the generally held belief that mood, policy ambiguity, and economic uncertainty all have a major impact on changes in asset values. This expanding corpus of work offers important insights into the intricate connection between investment patterns, uncertainty, and more general economic swings.

Despite these developments, there is still a significant lack of academic research explicitly examining the relationship between domestic and international policy and economic uncertainties and natural gas prices in the United States.

The Economic Policy Uncertainty index (EPU, Baker 2012b), which gauges uncertainty over economic policy risks, is used to quantify economic uncertainty. Additionally used are the US equities market-related uncertainty index, which gauges investor sentiment, and the CBOE VIX, which measures predicted volatility from S&P 500 options. The EPU, which is widely used in the literature, is a gauge of sentiment on economic uncertainty in addition to acting as a stand-in for the lack of agreement on economic policy. The VIX, sometimes known as the "fear index," is a crucial emotion indicator for the stock market since it increases during stressful times and decreases under favorable market conditions (Whaley 2009).

In our analysis, we use a two-covariate GARCH-MIDAS model, which is renowned for its exceptional predictive ability and ability to provide both short- and long-term volatility projections at the same time. According to Conrad and Kline (2020) and empirical experience, adding more than two covariates to the GARCH-MIDAS model at this point causes the program to crash. We study both short- and long-term volatility under various economic conditions. The US national financial condition index (NFCI) and US industrial production are two low-frequency factors that affect AIA's long-term volatility. According to earlier research by Dungey and Gajurel (2014) and Bekaert et al. (2014), the US financial and economic environment is the main cause of equities market contagion.

We use a two-step approach to solve the difficulty of estimating GARCH-MIDAS models with extra variables. We use quantile regressions (MSR and QR) and Markov switching to define the relationships between volatility and uncertainty indicators in both low- and high-volatility scenarios.

Our findings highlight the importance of sentiment and uncertainty indicators in explaining both short- and long-run volatilities, with a more substantial impact on long-term AIA volatility across varying market volatility regimes. This underscores the significance of VIX, EPU, GPECU, and GPCUCH in influencing volatility patterns. Notably, our evidence provides an intuitive explanatory variable for volatility clustering/persistence, challenging the conventional statistical treatment of these patterns.

The results endorse the GARCH-MIDAS approach, showcasing the relevance of economic certainty indicators like EPU and VIX, GEPU, and EPUCH in both high- and low-volatility regimes. These indicators serve as leading indicators, signaling changing sentiments leading up to abnormal volatilities. This observation is crucial for wealth managers, emphasizing the utility of volatility indices in predicting heightened volatilities in major Pacific Rim equity markets.

Our paper contributes significantly by jointly analyzing the linkage between short- and long-term volatility components with market sentiment and economic uncertainty. Additionally, we enhance the Su et al. (2019) model by deriving short- and long-term AIA volatility measures from a two-covariate GARCH-MIDAS model. The inclusion of major market uncertainty trackers in the Markov switching and quantile regressions addresses potential specification errors, providing a more robust analysis. These uncertainty trackers, shown to be significantly associated with short- and long-term volatility in asset returns, contribute to the methodological improvement of our study (see Pan et al. 2017, Su et al. 2017, 2021, among others).

The subsequent sections of the paper are structured as follows. In Section 2, we offer a concise review of the literature on economic uncertainty and, separately, the GARCH-MIDAS methodology employed in this study. Section 3 delves into the data and their respective sources, while Section 4 provides brief explanations of the methodologies applied. The empirical results are presented and discussed in Section 5, and Section 6 concludes the paper with a summary and some closing remarks.

2. Literature review of EPU and GARCH-MIDAS

This section provides a brief review of literature on the EPU and other uncertainty indicators that we deploy in this study. It also summarizes some of the pertinent studies that use GARCH-MIDAS framework for equity returns volatility. A complete review of literature can be found in Adrangi et al. (2023)

*2.1 Literature on Economic Policy Uncertainty*
Researchers have become more interested in studying how emotion and uncertainty indicators relate to the volatility of asset prices. Adrangi et al. (2023) go into great detail about the current body of research on this subject. Although we emphasize that this is not the main focus of the current research, we provide a modified version of their presentation in this section. Our goal is to frame and contextualize the current study. The global economic policy uncertainty index (GPECU), the China economic policy uncertainty index (GPUCH), and Twitter-derived measures of economic uncertainty (TEU) are a few of the indices that Baker et al. (2012a, 2012b, 2016, 2021, 2019) developed to capture economic and financial uncertainty. NVIX, a monthly news-based indicator, was first presented by Manela and Moreira (2017). Particularly when considering the volatility of different asset classes, these indicators have become more popular

(e.g., Pan et al. 2017, Su et al. 2017, 2018, 2019, Altig et al. 2020, Krol 2014, Dutta et al. 2021, Jiang et al. 2019, Xu et al. 2021, Pan et al. 2021, Lindblad 2017, Colak et al. 2017, Demir and Ersan 2017, Walkup 2016, Phan et al. 2019, Jens 2017. Gulen and Ion 2015, among others).

The relationship between the US and international equities and commodity markets and a number of indices that measure economic, market, financial, and policy risks has been thoroughly studied by researchers. Other indices utilized in this study have not gotten as much attention as the Economic Policy Uncertainty index (EPU), which has been used extensively. In order to quantify the degree of uncertainty around fiscal and monetary policies, Baker et al. (2012b) created the Economic Policy Uncertainty Index (EPU), which showed fluctuations over time and peaked in August 2011. The study discovered that weaker economic growth is linked to increased economic policy uncertainty, which affects investment, employment, and production.According to a review of previous studies on the relationship between EPU and financial markets as well as corporate behavior, firms typically adopt conservative strategies when EPU is high (Al-Thaqeb & Algharabali, 2019). Reducing capital investment, lowering the number of initial public offerings, limiting mergers and acquisitions, cutting dividend payments, and holding onto cash are all examples of this. The EPU seems to be a stand-in for uncertainty around company performance, stock market volatility, and economic policy.

Other researchers have looked into the relationship between realized volatility in equity markets and economic policy uncertainty (EPU), including Antonakakis et al. (2013), Liu and Zhang (2015), and Arouri et al. (2016). According to their findings, an increase in EPU lowers stock returns, especially when there is a lot of volatility. When taken as a whole, these studies highlight how important uncertainty indicators are to comprehending market dynamics and volatility trends.

The ability of the economic policy uncertainty (EPU) index to measure the risks of international contagion has been the subject of several studies. Tsai (2017) explores the impact of EPU on the contagion risk of investments across 22 international equity markets in China, Japan, Europe, and the US. According to the study, EPU has a major impact on market volatility in China and Europe, respectively. The interconnectivity of the equities markets is the reason for this correlation.

In a regime-switching setting, Choi and Hammoudeh (2010) examine conditional correlations (DCCs) between a number of assets, such as crude oil and the S&P 500. Using Markov-switching regressions, Chang (2022) investigates synchronous and asynchronous volatility relationships between the Japanese and American stock markets, taking into account the EPU index. The findings show that patterns are shifting from synchronous to asynchronous, with U.S. markets seeing the majority of EPU contagion to equities volatility. Together, these studies highlight how important the EPU index is for comprehending the dynamics of volatility and contagion in global financial markets.

Wen et al. (2019) investigate the connection between China's macroeconomic factors and the U.S. EPU. Other researchers have looked into the impact of the U.S. EPU on a number of financial markets, including bond markets (Liow et al. 2018), currency rates (Krol 2014), real-time economic uncertainty indexes (Altig et al. 2020), and crude oil prices (Sharif et al. 2020).

*2.2 Literature on GARCH-MIDAS and related applications*

This section summarizes some of the literature on the approaches we use in the current investigation, specifically GARCH-MIDAS, quantile regressions (QR), and Markov Switching regressions (MS), in accordance with Adrangi et al. (2023). The primary objective is to demonstrate how the GARCH approach has continued to advance and evolve, most recently into the GARCH-MIDAS frameworks. Our study builds on the work of Su (2017) and Pan (2021) and follows in their footsteps.

Significant progress has been made in ongoing efforts to improve forecasting and volatility measuring techniques, especially within the GARCH framework. The groundbreaking work of Engle and Lee (1999) and Ding and Granger (1996) has influenced further studies and demonstrated improvements in GARCH model estimates. The validity of these techniques is further supported by Golchin and Rekabdar (2024).

A GARCH (1,1) model with a convexity requirement is proposed by Javaheri et al. (2004) to study the hedging of volatility swaps. According to Awartani and Corradi's (2005) evaluation of the out-of-sample volatility prediction capabilities of different GARCH model variants, asymmetric GARCH models outperform GARCH for longer-horizon and one-step-ahead forecasts (1,1). Performance tests by Liu and Hung (2010) show that the GJR-GARCH model produces the best accurate volatility projections for the S&P 500, with the EGARCH model coming in second. To improve volatility forecasting for the S&P 500, Hajizadeh et al. (2012) provide hybrid models that combine EGARCH with artificial neural networks. Robust substitute techniques are proposed by Carnero et al. (2012) to enhance GARCH volatility estimations and mitigate possible upward estimation bias. Adrangi et al. (2015) investigate volatility spillovers between the crude oil and equities markets by estimating bivariate vector autoregressive EGARCH models.

Over the last ten years, adding data with different frequencies has been an innovative way to improve the performance of GARCH models. The mixed frequency data sampling technique in conjunction with GARCH-MIDAS modeling was first presented by Engle et al. (2013). The GARCH-MIDAS model has since been used in studies by Asgharian et al. (2013), Conrad and Loch (2015), Lindblad (2017), Amendola et al. (2017, 2019), Pan et al. (2017, 2021), Conrad et al. (2018), and Borup and Jakobsen (2019), Amado (2019). These studies show that accuracy is increased when low-frequency macroeconomic data are included in high-frequency volatility projections.

The relationship between WTI and Brent crude oil spot price volatility and market fundamentals is examined by Pan et al. (2017) using a regime-switching univariate GARCH-MIDAS model, which produces better crude oil volatility forecasts. Fang et al. (2018) demonstrate how shifts in U.S. investor confidence affect G7 equity markets using a two-covariate GARCH-MIDAS model. Zhu et al. (2019) use a uni-covariate GARCH-MIDAS model to study daily volatility in key stock indexes and look at equity market volatility indices based on text counts of newspaper articles. A GARCH-MIDAS method is used by Conrad et al. (2018) to determine the short- and long-term volatility components of cryptocurrencies. When taken as a whole, these studies demonstrate the development of GARCH models and their uses, emphasizing the advantages of using the GARCH-MIDAS technique.

Su et al. (2019) examine the impact of uncertainty on the short- and long-term volatility of stocks in nine economies, which is of special interest to this study. They break down the conditional variance into short-term and long-term components using a two-variable GARCH-MIDAS model. The daily volatility is the short-term component. They estimate a linear function for low frequency measurements of uncertainty for the long-term component. They take into account the EPU, FU, and NVIX, three indicators of US market uncertainty. The three categories of fundamental components make up the EPU. Newspaper coverage of economic uncertainty connected to policy is measured by one component. The number of federal tax code provisions that are scheduled to expire during the next ten years is reflected in the second component. The disagreement among economic forecasts serves as a stand-in for uncertainty in the third component. A common component in the time-varying volatilities of h-step ahead forecast errors across a wide range of financial indicators is measured by FU, which was proposed by Ludvigson et al. (2021).

According to Su et al. (2019), there is a positive correlation between industrialized nations' equity market volatility and Economic Policy Uncertainty (EPU). Surprisingly, reduced volatility is associated with a higher National Financial Volatility Index (NVIX) in terms of forecasting power. The model's prediction ability is not greatly enhanced by the Financial Uncertainty Index (FU), especially when it comes to long-term stock market volatility. Despite their conflicting findings, they add important new information to the body of knowledge regarding the critical role uncertainty indexes play in the volatility of the stock market. Notably, Su et al. (2019) did not incorporate all three uncertainty indicators into their two-covariate GARCH-MIDAS model, presumably in order to circumvent difficult estimate problems. But by restricting the model to two of the three uncertainty trackers, this modeling decision may create misspecification, endangering accuracy and producing erroneous results.

Pan et al. (2021), on the other hand, suggest a collection of univariate GARCH-MIDAS models for Expected Shortfall and Value at Risk (VaR). By minimizing the loss function proposed by Fissler and Ziegel (2016), they determine the model parameters. Their models, which use the S&P 500 daily returns as well as the producer price index (PPI) and industrial production (IP) monthly values, show that macroeconomic uncertainty has a major impact on the long-term volatility component. The univariate character of Pan et al.'s (2021) model, which addresses each volatility metric separately, is a drawback, though, and raises questions regarding possible misspecification. Furthermore, they don't contrast their models with other GARCH-MIDAS models or bivariate models.

Quantile regressions have been used in a number of research to investigate the connection between sentiment indicators and volatility, such as Dutta et al. (2021) and Xu et al. (2021). The relationship between news-based Equity Market Volatility (EMV) and crude oil volatility is examined by Dutta et al. (2021) using quantile regressions, and they find a significant but asymmetric influence at times of high oil volatility. A univariate GARCH-MIDAS (QR-GM) model based on quantile regression is used by Xu et al. (2021) to forecast Value-at-Risk (VaR) in daily spot and futures returns for crude oil. The effectiveness of GM and QR-GARCH-MIDAS models in examining how low-frequency factors affect the quantile of high-frequency dependent variables is demonstrated by their findings. However, the fact that Xu et al. (2021) only used one covariate, GEPU, in the GM model, is a disadvantage of their study.

Conrad and Kleen (2020) show that the multiplicative GARCH-MIDAS model performs better than several GARCH model variations, such as the MS-GARCH, the high-frequency-based volatility (HEAVY) of Shephard and Sheppard (2010), the realized GARCH of Hansen et al. (2012), and the heterogeneous autoregression (HAR) of Corsi (2009). Their findings on the GARCH-MIDAS method's superior performance encourages Adrangi et al. (2023) and the current study to embrace it.

The wealth of research on volatility assessment and methods for examining its determinants indicates that investors, policymakers, fund managers, and speculators are interested in the topic.

Our work is an extension of the earlier body of work on the topic. Before this study, a two-variable GARCH-MIDAS model was also used by Fang et al. (2018), Su et al. (2019), and most recently, Adrangi et al. (2023) to examine the relationship between market uncertainty and stock market volatility. Nevertheless, Conrad and Kleen (2020) propose that "GARCH-MIDAS models with more than two variables in the long-term component are challenging to estimate due to the likelihood being relatively insensitive with respect to changes in the weighting parameters." Based on our real-world experience, it is impossible to estimate GARCH-MIDAS models for modeling long-term volatility that include more than two covariates. The main cause of the widespread use of univariate GARCH-MIDAS models in earlier studies is this restriction, which makes them vulnerable to misspecification. To make estimate easier, Su et al. (2017) and Pan et al. (2017) incorporate several combinations of uncertainty measure pairs into their GARCH-MIDAS (GM) models. Although this approach makes estimation feasible, it is still lacking since it does not account for certain uncertainty metrics.

By incorporating economic principles into the GARCH-MIDAS model, our research adds to the body of current knowledge. Furthermore, in a parallel second step, we add uncertainty measures to Markov Switching (MS) and Quantile Regression (QR) regressions. This methodological innovation enables us to investigate the association between several variables and market volatility in two steps. Therefore, using a multivariate framework, our method investigates the relationship between the uncertainty variables and the volatility of AIA over both the short and long term.

3. Data

We sample data between November 21, 2007 and December 31, 2023 to derive volatility estimates in AIA. Monthly data on the national financial confidence index (NFCI) and US industrial production are taken from the Federal Reserve Bank of St. Louis (FRED).

The daily data for iShares Asia 50, AIA for the same time frame come from Yahoo Finance. AIA is selected because it tracks the investment results of the S&P Asia 50 composed of 50 of the largest Asian equities, in all key sectors of major Asian economies such as Samsung, Alibaba, Taiwan Semiconductor, among others. The fund invests at least 80% of its assets in the component securities of its index or similar securities and may invest up to 20% of its assets in certain futures, options and swap contracts, cash and cash equivalents. We are not considering individual equities because studying individual equities adds company level attributes such as managements, business practices, financial ratios to the model which could result in complexities and distracts from the aim of the paper. Furthermore, many of the studies of the subject of economic uncertainties also consider indices rather than individual equities. The daily index values of the VIX and the EPU in the US are taken from the Bloomberg and the FRED (St. Louis) databases, respectively.

The daily news-based EPU index is constructed using the archives of Access World News's News Bank service. This global database stores archives of thousands of newspapers and other news sources. The EPU index is based on more than a thousand US newspapers, ranging from large national papers, such as *USA Today*, to small local ones. The index is computed by determining the number of newspaper articles that contain the words *economy*, *uncertainty*, *legislation*, *deficit*, *regulation*, *Federal Reserve*, or *White House*. This number is then normalized by the total number of newspaper articles. The EPU index is updated daily for the current and past months.

We also include the daily index of the leading indicator of implied market volatility, the VIX. The Chicago Board Options Exchange (CBOE) introduced the first volatility index in 1993, which was known as the VXO. It was based on implied volatilities from at-the-money options on the S&P 100 index, using a methodology proposed by Whaley (1993). The CBOE used an alternative methodology in 2003 to calculate the VIX as a weighted sum of out-of-money option prices for all S&P 500 strikes in real time. Whaley (2009) discusses the public and media interest in the value of the VIX as a measure of volatility and explains the origin and purpose of creating the VIX and its role in explaining the state of the economy and equity markets.

The VIX is intended to measure the expected price fluctuations in the S&P 500 index options over the next 30 days. Market participants have used the VIX and its predecessor, the VOX, to gauge the market sentiment in the US and around the world. We use the VIX as an indicator of the future financial market risk because the financial press quotes the VIX volatility index as a gauge of investor fear. Governmental agencies and central banks use the VIX to assess risk in financial markets. Previous research has shown that, though far from perfect, the VIX does have some forecasting power. Moreover, a strong association exists between the VIX and contemporaneous price dynamics: positive shocks to the VIX are associated with declining markets and vice versa. It follows, therefore, that an elevated VIX portends weaker prices in the future. In that sense, the VIX captures both fear, and price dynamics – a high VIX indicates fear associated with market declines (e.g., Whaley (2009)). While VIX and EPU and other model variables like GEPU, ECU and EPUCH may appear strongly correlated, the VIFs from estimated equation (1) are unanimously below two indicating that there are no serious collinearities among the model explanatory variables. These results are not presented but are available from authors.

The equity market-related economic uncertainty index (ECU) created by Baker et. al (2012a, 2012b), is another appropriate index to that is potentially informative as a leading indicator of the US equity markets turmoil. Baker et. al (2016) constructed the monthly Global Economic Policy Uncertainty Index (GPEU) as a GDP-weighted average of national EPU indices for twenty economies of the world, including the US. The monthly Chinese Mainland EPU newspaper-based indices of policy uncertainty in China is based on the work by Davis et. al (2019). The index quantifies uncertainty-related concepts from October 1949 onwards using two mainland Chinese newspapers: the Renmin Daily and the Guangming Daily. Quantile and Markov Switching Regressions are also based on mixed frequency data that includes daily and monthly observations.

4. Methodology

In this section, we provide a concise overview of the empirical methodology employed in this research, closely aligned with that of Adrangi et al. (2023). It is worth to note that the detailed explanation of the methodology is not the focal point of this research and is presented solely for the benefit of the readers. Interested readers may review the original work on the subject by Conrad and Kleen (2020).

We deploy the multiplicative GARCH-MIDAS methodology, to derive short- and long-term volatility estimates. However, GARCH-MIDAS estimation tends to crash as indicated by Conrad and Kleen (2020) and our own experience if more than two covariates are included in the estimation. Therefore, we adopt a two-stage approach. In the first stage we drive the short- and long-term volatilities that stem from the US equity markets. Daily AIA return series, monthly NCFI and industrial production are the high- and low-frequency series included in the GARCH-MIDAS estimation of short- and long-run volatilities in AIA (STV, and LTV, respectively). The US NCFI is intended to gauge the investment sentiments while the industrial production aims to account for the health of the US economy and the potential demand for equities. It has been shown that the US equity market conditions tend to spill over through contagion to the equity markets around the world (see Boubaker et al. (2016)). In a similar study, Engle et al. (2013) use monthly industrial production growth and monthly inflation as explanatory variables.

Having obtained estimates for STV and LTV, we estimate a set of regressions that summarize the relationship between the dependent variable (i.e., STV or LTV) and the explanatory variables. Equation (1) is the implicit regression model.

$$\text{AIA volatility} = f(\text{ECU, EPU, EPUCH, GEPU, VIX}) \qquad (1)$$

Before testing the association of AIA return volatility with the potential explanatory variables, we examine the model variables for possible structural breaks. If we find breaks in the series, we interpret them as periods of regime change. AIA Volatility in the short-and long-term will plausibly vary during these regime changes. The Markov switching and Quantile regression are well equipped to capture changes in the state of volatility. For instance, quantile regressions appropriately examine the association of uncertainty indices with volatility at low to high quantiles of the volatility of AIA. Similarly, Markov switching regression show the manner of transition of these relationships from one regime to another. Furthermore, it is necessary to test each time series for stationarity. Visual inspection shows that all model variables appear stationary, but not surprisingly, STV and LTV exhibit possible structural breaks which could occur due several events including COVDI19 pandemic. We address these issues next.

*4.1. Structural Breaks and Stationarity*

We apply Bai and Perron's (2003) test of structural breaks and examine the stationarity of the series under study by deploying the augmented Dickey–Fuller (ADF; Dickey and Fuller (1979)) and the Phillips–Perron (PP; Perron and Phillips (1988)) tests. Tests of structural breaks and stationarity are well covered by the literature; in the interest of brevity, we do not explain them in detail here.

*4.2. GARCH-MIDAS*

Adrangi et al. (2023) provide a succinct summary of the GARCH-MIDAS methodology. We adhere to their explanation to familiarize the reader with this approach, recognizing that it is not the primary focus of this research. Bollerslev (1986) introduced the GARCH model as an enhancement to conventional autoregressive conditional heteroskedasticity (ARCH) time-series models, aiming to better capture the observed volatility clustering in financial markets. Unlike the assumption of homoscedasticity in a typical ARCH model, the GARCH model accounts for temporal heteroskedasticity, a characteristic often present in financial market data. Volatility clustering is evident in both prices and rates of return, where periods of high volatility tend to be followed by low volatility and vice versa. The variability in financial markets may exhibit irregular patterns, contributing to heteroskedasticity. Notably, financial markets often experience heightened volatility during actual or perceived financial crises, contrasting with periods of calmness during phases of steady economic growth.

We estimate the GARCH-MIDAS based on MIDAS regression models, which were introduced by Ghysels et al. (2004, 2016) to generate the short- and long-term real volatility in the daily returns of the AIA. Our two covariate GARCH-MIDAS model includes monthly values of the NFCI and the changes in the industrial production in this study. MIDAS methodology offers a framework to incorporate variables of different frequencies to obtain multi-horizon volatility. As shown by Adrangi et al. (2023) the MIDAS regression model is expressed as equation (2):

$$y_{t+k} = \alpha_0 + \alpha_1 x^m_t + \varepsilon^m_t, \qquad (2)$$

where $y_{t+k}$ is the $k$ step-ahead value of the dependent variable at time $t$ with the highest frequency, $x^m_t$ may be a vector of independent variables at time $t$ and $m$ is the frequency matching that of y, $\varepsilon^m_t$ is the random innovation at time $t$ with $m$ frequency, and $\alpha_0, \alpha_1$ are the intercept and a conformable vector of model coefficients. The expressions for $\varepsilon_t$ and its variance in GARCH (1,1) specification are:

$$\varepsilon_t = (\varepsilon_{t-j})\sqrt{\sigma_{\varepsilon,t}^2}$$

and

$\sigma_{\varepsilon,t}^2 = f(\varepsilon_{t-1}, \sigma_{t-1}^2)$, for GARCH (1,1) specification.

Similar to the work of Engle et al. (2013), Ghysels et al. (2016), and Conrad and Kleen (2020), we combine the high-frequency daily data with low-frequency monthly data in the GARCH-

MIDAS model. The conditional variance of innovations is decomposed into short- and long-term volatility components multiplicatively as:

$$\sigma^2_{\varepsilon,t} = h_{i,t}\tau_t, \tag{3}$$

where $h$ and $\tau$ capture the short-term volatility of the high-frequency data and the long-term volatility, respectively, given by equations (4) and (6).

The short-term component of GARCH-MIDAS in equation (3) is taken from Engle et al. (2013) which uses the GARCH process of Bollerslev (1986).

$$h_{i,t} = a_0 + a_1 \frac{\varepsilon^2_{t-1}}{\tau_t} + a_2 h_{i-1,t}. \tag{4}$$

The short-term volatility component $h_{it}$ varies daily within period t that signifies the high frequency. It is designed to capture daily volatility clustering and is a mean-reverting unit-variance GJR-GARCH (1,1) process. We re-write equation (4) to break down each coefficient into its estimated components as equation (5).

$$h_{i,t} = (1-\alpha-\gamma/2-\beta) + (\alpha+\gamma|_{\varepsilon_{i-1,t}<0})\frac{\varepsilon^2_{t-1}}{\tau_t} + \beta h_{i-1,t}. \tag{5}$$

Equation (6) is the basis for the long-term volatility for which the realized volatility is smoothed over K periods, and will be expanded by adding low frequency covariates.

$$\tau_t = m + \theta \sum_{k=1}^{K} \phi_k(w_1, w_2) x_{t-k}, \tag{6}$$

The short-term component accounts for the well-recorded volatility clustering in the daily AIA returns. The long-term component is constant across days and changes at lower frequency, consistent with low frequency series in the GARCH-MIDAS model, which in this paper is bi-monthly consistent with the frequency of the NFCI included in the GARCH-MIDAS model, similar to Conrad and Kleen (2020) specification. Equation (6) may be further enhanced to include other variables and their variances.

The weighting scheme in equation (6) is given by

$$\phi_k(w_1, w_2) = \frac{(k/K)^{w_1-1}(1-k/K)^{w_2-1}}{\sum_{j=1}^{K}(j/K)^{w_1-1}\sum_{j=1}^{K}(1-j/K)^{w_2-1}},$$

Where

$$\sum_{k}^{K}\phi_k(w_1, w_2) = 1.$$

The weighting scheme in (7) generates hump-shaped or convex weights. As $w_1$ is restricted to 1 (see Su et al. (2017), Fang (2018), Conrad and Kleen (2020), the weighting scheme guarantees a decay pattern where the rate of decaying is determined by parameter $W_2$. The restricted weighting scheme boils down to

$$\phi_k(w_2) = \frac{(1-k/K)^{w_2-1}}{\sum_{j=1}^{K}(1-j/K)^{w_2-1}}.$$

*4.3. Regime Switching Markov Regression*

In a framework involving regime switching, a variable y might be influenced by a discrete state variable that remains unobservable. This regression model is suitable when there is an assumption of multiple regimes in the underlying data-generating process. At any given time t, the process could be situated in state $S_t$. The switching model permits the use of distinct regression models for each regime. The conditional mean of yt in regime m, given a set of switching regressors x and corresponding coefficients, along with non-switching vectors of regressors zt and coefficients, is expressed by equation (7). Adrangi et al. (2023) provide an in-depth explanation of the methodology. Hence, we present a condensed version of their exposition.

$$\mu_t^{(m)} = x_t'\beta_m + z_t'\phi. \tag{7}$$

Assuming that the regression error $\varepsilon_t$ is independently and identically distributed (iid) and its variance may be regime dependent, the regime switching regression model may be expressed by equation (8).

$$y_t = x_t'\beta_m + z_t'\phi + \sigma_m \varepsilon_t. \tag{8}$$

Regime probabilities may be assumed to be a function of a vector of exogenous variables and parameters. The multinomial logit expression of the regime probabilities is given by equation (9).

$$P(s_t = m | \Omega_{t-1}, \psi) = p_m(E_{t-1}, \psi) = \frac{\exp(E'_{t-1}, \psi_m)}{\sum_{j=1}^{M} \exp(E'_{t-1}, \psi_j)} \tag{9}$$

In equation (9),

$m$ is a given regime,

$p_m$ is the regime probabilities,

$\Omega_{t-1}$ is the information set at time $t$-1,

$E_{t-1}$ is the vector of exogenous observable variables, and

$\psi$ represents model coefficients.

The model parameters are estimated through iterative optimization of the log of the likelihood function, based on equations (8) and (9), and given as equation (10).

$$l(\beta, \phi, \sigma, \psi) = \sum_{t}^{T} \log \left\{ \sum_{m=1}^{m} \frac{1}{\sigma_m} \varpi[\frac{(y_t - \mu_t(m))}{\sigma_m}].P(s_t = m | \Omega_{t-1}, \psi) \right\}. \tag{10}$$

The likelihood function in equation (10) is maximized with respect to the parameters $\beta, \phi, \sigma, \psi$ using iterative nonlinear optimization methods. The regime probabilities are derived by filtering while optimizing equation (10).

The transition matrix for $M$ regimes may be written as equation (11),

$$p(t) = \begin{pmatrix} p_{11}^t & \cdots & a_{1M}^t \\ \vdots & \ddots & \vdots \\ a_{M1}^t & \cdots & a_{Mn}^t \end{pmatrix}. \tag{11}$$

Each row of the transition matrix in equation (11) is defined by a distinct multinomial logit. By computing the one-step-ahead predictions of the regime probabilities and the Markov transition matrix, one may arrive at the one-step-ahead joint densities of the data and regimes in period *t*.

*4.4. Quantile Regression*

The extreme movements and structural breaks in the VIX and EPU series potentially have asymmetric effects on AIA volatility in the short and long term. Market participants and investors are not only sensitive to the smoothed association of the VIX, EPU, ECU, GEPU, EPUCH with AIA volatility; they are also interested in the impact of extreme up-and-down movements of the uncertainty indices and their association with AIA volatility. Structural breaks may be one source of extreme fluctuations in volatility.

As Adrangi et al. (2023) discuss, quantile regressions (QR) are well designed to capture the asymmetric dependence between dependent and explanatory variables. Furthermore, we are seeking to examine the robustness of the findings of MSR estimations. Specifically, our aim is to find out if the empirical findings from the MSR are bolstered or corroborated by the QR approach in any quantiles.

We follow Adrangi et al. (2023) and offer a brief summary of the QR methodology in this paper. QR models are nonlinear (see e.g., Galvao et al. 2020) and robust in the presence of extreme events and asymmetric dependence when the assumption of linearity may not be appropriate (see Geraci 2019; Yu et al. 2003, Hendricks & Koenker 1992). They are superior to OLS estimates because they allow coefficient estimates to vary with the distribution of the dependent variable, thus, accurately modeling the relationship between the explanatory variables and the dependent variable. Following is a brief explanation of QR.

Suppose that we have a random variable *Y* (AIA volatility) with probability distribution function

$$F(y) = \text{Prob}(Y \leq y)$$

so that, for $0 < \tau < 1$, the $\tau$th quantile of *Y* may be defined as the smallest *y* satisfying

$$F(y) \geq \tau:$$

$$Q(\tau) = \inf\{y : F(y) \geq \tau\}.$$

The empirical distribution function is given by

$$F_n(y) = \sum_s l(Y_i \leq y),$$

where $l(.)$ is a binary function that takes the value 1 if $Y_i \leq y$ is true and 0 otherwise. The resulting empirical quantile is given by

$$Q_n(\tau) = \inf\{y : F(y) \geq \tau\}.$$

Alternatively, the empirical quantile may be expressed as an optimization problem as:

$$Q_n(\tau) = \arg\min\left\{\sum_i \rho_\tau(Y_i - \omega)\right\},$$

where $\rho_\tau(w) = w(\tau - 1(w < 0))$, which asymmetrically assigns weights to positive and negative values in the estimation process.

Allowing for regressors X and assuming a linear specification for the conditional quantile of the dependent variable AIA volatility given values for the vector of explanatory variables leads to QR as

$$Q(\tau \mid X_i, \beta(\tau)) = X_i'\beta(\tau), \tag{12}$$

where in equation (12) $\beta(\tau)$ is the vector of coefficients associated with the $\tau$ th quantile.

The conditional quantile regression estimator is

$$\beta_n(\tau) = \arg\min_{\beta(\tau)}\left\{\sum_i \rho_\tau(Y_i - X_i'\beta(\tau))\right\}.$$

The quantile regression estimator is derived as the solution to a linear programming problem. We use a modified version of the Koenker and D'Orey (1987) version of the Barrodale and Roberts (1973) simplex algorithm.

The Powell (1986) kernel approach computes a kernel density estimator, using the residuals of the original fitted model,

$$\hat{H} = 1/n \sum b_n^{-1} K(\vartheta(\tau)/b_n X_i X_i',$$

where $K$ is a kernel function that integrates to 1 and $b_n$ is a kernel bandwidth. For bandwidth specification, we employ a method suggested by Hall and Sheather (1988) and a kernel bandwidth suggested by Koenker and Ng (2005). Koenker and Machado (1999) define a pseudo R-squared for the goodness-of-fit statistic for quantile regression that is analogous to the R-squared from conventional regression analysis.

5. Empirical Findings

*5.1. Structural Breaks and Stationarity*

Figures 1 and 2 present the grpah of STV and LTV, respectively. The time seris plots suggest that there are posibilities of structural breaks and LTV could be non-stationary. The Bai–Perron test signals four structural breaks in the short- and long-term volatility in the AIA that occurred on several occasions. The break dates in the STV do not coincide with those of the LTV and could be due to many events. The year 2012 witnessed geopolitical and economic upheavals and possibly impacted global equity markets immediately or with some time lag is it usually is the case.  For instance, wars in Syria and Afghanistan continued to rage.  The European Union announced that it was imposing oil embargo on Iran.   These without a doubt shocked the crude oil markets for a few months before a new equilibrium was achieved.  Similarly, world equity markets absorbed other shocks from ISIS success in Iraq which continued to threaten crude oil supplies from that country.  Taliban were also expanding their operations with destabilizing political and economic effects in the entire region.

Breaks in 2014 were possibly related to failed nuclear negotiations with Iran, crude oil price crash, faltering EU economies.   Similarly, 2016 was market by significant economic and geopolitical evens.   The Trans-Pacific Partnership (TPP) was unraveling under the administration of the US President Donald Trump.  This had taken seven years of hard negotiations and was the largest regional trade pact for the US and its trading partners in East Asia in history.  The eleven nations who had joined the US in this treaty, were forced to rethink the future of their trade with the US and possibly entering negotiations with the second largest economy of the world, china. Britain and Whales voted to leave the EU while Scotland and Ireland opposed this move. Markets expected the Brexit saga to continue and the fallouts on UK EU economies remained uncertain. North Korea also intensified it missile tests in response to the US saber rattling. Effects of these events on world trade, security, and economies were unsettling. Among the events of 2019 that fueled market with uncertainty and rocked the investment world are nuclear impasse between the US and North Korea, the Brexit are the most salient.

These evens cast pall on global trade, fueled inflation worries, and contributed to risk and uncertainty for equity markets of the world. The year 2021 was gripped with COVID19-related supply chain disruptions, shocks to many firms and industries. Fears of persistent inflation kept equity and commodity markets volatile.

Examining the graphs of the short- and long-term AIA volatility and the uncertainty indicies under study, suggests possible mean and covariance stationarity for most variables with the exception of he LTV. To confirm the graphic evidence (not presented, for the purpose of brevity), we conducted formal statistical tests. Table 1, Panel B presents the statistical evidence of the behavior of these series. As shown, all variables under study are found to be stationary, employing the ADF tests with the inclusion of the break points.

*5.2 Markov Switching Regression Results*

Table 2 shows the coefficient estimates of the GARCH-MIDAS model that produces the short-term and long-term volatility components of the realized volatility in AIA. Short- and long-term volatilities formulated by equations (5) and (6), respectively are estimated simultaneously. The estimated $\beta$ of 0.896 is statistically significant suggesting strong persistence in the short-term volatility component. Su et al. (2017) find similar results in their research. The statistically significant $\alpha$ and $\Upsilon$ coefficients in equation (5) are also plausible and confirm the validity of equation (4) in the GARCH-MIDAS estimation. The $\theta$ coefficients represent the changes in the long-term volatility stemming from the lagged effects of the industrial production and NFCI, respectively. The coefficient θ is statistically significant and negative, indicating that a rise in the US industrial production reduces the long-term volatility in AIA as expected. The rise in NCFI and the long-term volatility are negatively associated as shown by the negative and statistically significant $θ\_2$, which is plausible. A rise in the US financial confidence index represents improving investment climate in the US and thus, plausibly lead to a fall in the long-term volatility component of AIA. These findings also suggest the sensitivity of the Asian blue chip equities to the US economy. Trade with the US is a significant opportunity from companies like Samsung, Alibaba and Taiwan semiconductor, among others.

Table 3 presents the estimation results of the MSR. Examining the STV, we find that it rises in response to the US and China policy uncertainties and VIX during low volatility regimes. Thus, investors in AIA are focused on the policy changes of the two major world economies, as well as financial market indicators like VIX. However, these associations change under high volatility regime (regime 1). Under regime 1, STV is either unaffected by all uncertainty indicators or doesn't rise, with the exception of global policy uncertainties. The interpretation is that under conditions of high market volatility AIA investors are desensitized to any marginal uncertainties under study.

The US ECU, China policy uncertainties, and VIX raise long-term volatility in AIA across both low (regime 2) and high volatility (regime1) regimes. The US and global EPU appear to reduce the long-term volatility. We interpret these findings as AIA investors' long-term concerns are focused on market conditions in the US contained in VIX and ECU and economic policies of China. It is plausible that this finding is confirming that China EPU rather than the US and global policy uncertainties drive AIA long-term volatilities. The global and US economic policies do not contribute to the long-term volatility in AIA. These findings are supporting the notion that long-term investors in blue chip Asian equities are sensitive to policy changes in China, US economy (ECU) and the US equity markets (VIX). These observations also emphasize the economic and financial market contagion between the US and Asian equity markets.

We also conclude that the attitude of market participants regarding the VIX and the EPU is dependent on the volatility regime and varies asymmetrically in the short-run but not so in the long-run. Lin and Zhang (2015), Su et al. (2019), and Tsai (2017) and Chang et al. (2022) also find a positive association between EPU and volatility in equity markets and EPU and exchange rates.

Table 4 summarizes the transition probabilities from low to high volatility regimes for both short- and long-term volatilities in the AIA. It appears that the probabilities of transitioning from high to low volatility or in reverse are quite low. The AIA experiences longer periods of short-term volatility during the high volatility regime 1 than during the low volatility regime 2. Long-term volatility durations under both regimes are not as dramatically different under two volatility regimes 1 and 2. The duration of the long-term volatility under low volatility regime higher that under high volatility regime. The interpretation is that long-term volatility in AIA is not significantly influenced by regime changes. One interpretation is that long-term volatilities in AIA are dependent on other fundamental forces rather than shifts in volatility regimes that inevitably occur. These findings are helpful in formulating hedging strategies for long-term investors in AIA equities. However, derivative and other hedging instruments are not well-developed in Asian equity markets.

Figures 3 and 4 plot the predicted one-step transition probabilities of being in low- and high-volatility regimes for the short- and long-term volatilities. The constant transition probabilities and the expected duration of each regime for both volatility cases are presented in Table 4. It is evident that the low-volatility regimes are expected to last longer than the high-volatility regime. Therefore, policy uncertainties and the VIX will be critical more often than not, which is plausible and indicates that equity markets do not favorably perceive uncertainties stemming from policy decisions or financial turbulence.

*5.3 Quantile Regression Results*

The QR estimation results of equation (1) are presented in Tables 5 and 6 for the short- and long-term volatility estimates of the AIA. The Lagrange Multiplier test for ARCH effects in Table 1 show the presence of ARCH effects in the STV but not LTV. The QR estimates remain robust and reliable in the presence of ARCH effects. Thus, the QR estimates of the coefficients at various quantiles of distribution of the AIA volatility are better suited to capture the relationship among the variables in the presence of heteroscedasticity and at extreme levels of the distribution as well as the conditional median.

Somewhat consistent with the MSR findings, the US ECU, and various policy uncertainties raise the short-term volatility in AIA at most quantiles. For instance, in most quantiles of STV, the associations of the STV with these measures of uncertainty are statistically significant, and have the expected positive sign. However, STV is not sensitive to VIX in several quantiles. That may change depending on the AIA investors' concerns. While Bekaert et al. (2014) and Dungey and Gajurel (2014) show that contagion among equity markets is common, investors my put more weight on the US, China, or global policy uncertainties. There is research in support of this phenomenon. Edwards et al. (1995) analyze media coverage of issues in the context of elections. They confirm that the public perception of salient issues varies over time. Judging by the coefficient signs and statistical significance, the US economic uncertainties and China policy uncertainties are crucial in the minds of AIA investors in the short-run and raise volatility in this index.

These OLS estimation results, which predict the association of the variables based on the conditional mean of the AIA volatility for both the STV and the LTV, may be misleading. For instance, Tables 5, the OLS estimation results show that there is no distinction among the uncertainty indices all of them raise short-term volatility in AIA.

Turning to the LTV, QR estimate results indicate that LTV in AIA is consistently responsive to the US ECU and China policy uncertainties, as well as VIX in most quantiles and rise in response to them. However, the association between the LTV and other measures of uncertainties in the study are erratic and statistically unstable. Thus, we conclude that AIA investors are responsive to China's policies and the US economic and equity markets uncertainties, ECU and VIX that may be spilling into Asian markets. AIA long-term volatilities appear to react erratically, or even decline in reaction to the global and the US policy uncertainties. Thus, the US and global economic policy uncertainties are not salient considerations for AIA investors. In summary, our findings lend support to contagion effects from the US economy and confirm that economic uncertainties in the US markets play a significant role for investors in the Asian blue chip firms. Furthermore, long-term volatilities in

AIA react to economic and equity market conditions in the US and not to US or global economic policy uncertainties.

The OLS estimation results, also corroborate the LTV positive reactions to the ECU, GPUCH and VIX. However, the conditional mean of the long-term volatility in AIA declines in reaction to the US and Global economic policy uncertainties confirming the notion that long-term volatilities in AIA are not associated with the US or global policy matters.

The main take away is that STV and LTV react to policy uncertainties as well as economic and equity markets uncertainties in the US, represented by ECU and VIX. However, movements in STV and LTV are sensitive to the saliency of global or regional events in the minds of investors. For instance, if investors are focused on the US economy, then economic policy uncertainties in the US raise volatilities either in the short-term or long-term or both. China economic policies in both regimes appear to raise STV and LTV almost in all cases. The global policy uncertainties seem to play much smaller role in the volatility of AIA both in the short- and long-terms. In some cases it is even negatively associated with AIA volatilities.

Figures 5and 6 show the coefficient evolution process through quantiles for the STV and LTV. They show the responses of the short and long-term volatilities in every quantile to each measure of uncertainty holding all else constant. Figure 2 shows that the association of STV with all uncertainties measures is positive and statistically significant in most quantiles. This observation supports the findings of the MSR estimates and shows that these associations are dynamic and that coefficient signs and statistical significance are regime dependent. Figure 6 confirms that LTV is positively and statistically significantly associated with ECU, VIX and EPUCH consistent with reported findings in Table 6. In summary, findings of QR corroborate the MSR and confirm the robustness of the findings in this paper.

6. Summary and Implications

This study explores the association between equity market volatility in the Pacific Basin, as indicated by AIA volatility, and a spectrum of economic, financial fundamentals, and variables reflecting market and government policy uncertainties.

The equity markets of the Pacific Basin hold significant importance for both the United States and global financial markets due to their substantial economic weight, interconnectedness, and the role of the region as a key driver of global economic growth. The Pacific Basin encompasses major economies such as Japan, China, Australia, and emerging markets in Southeast Asia. As a collective force, these markets contribute significantly to the global GDP and serve as crucial trade partners for the United States. Investors and financial institutions worldwide are heavily invested in Pacific Basin equities, and movements in these markets can have ripple effects, influencing global asset prices, portfolio diversification strategies, and international capital flows. Moreover, the region's economic policies, geopolitical developments, and market trends can impact the stability and performance of global financial markets, making a thorough understanding of the Pacific Basin's equity dynamics imperative for policymakers, investors, and financial analysts across the globe.

A crucial aspect of our analysis involves investigating the combined impact of these uncertainties on both long- and short-term volatility patterns. Our dataset spans from 2007 to 2023. Utilizing a two-covariate GARCH-MIDAS model for equity market volatility estimation, we leverage its demonstrated superiority over other GARCH model variations in predictive accuracy and concurrent delivery of short- and long-term volatility predictions. The short- and long-term AIA volatility estimates are analyzed for their responses to equity market uncertainties, uncertainties in the US economic policies, global and China economic policy uncertainties, and implied risk measured by the CBOE VIX amid market regime changes. Employing Markov switching and quantile regressions facilitates a nuanced exploration of the relationship between volatility and uncertainty indicators during both low- and high-volatility periods. Furthermore, unlike previous studies, our MSR and QR multiple regressions on realized volatility include all the uncertainty variables simultaneously. Previous research primarily relies on one or two covariate GARCH-MIADS models, which face estimation problems when incorporating more than two covariates simultaneously.

Our analysis, conducted through MSR and quantile QR, reveals that the relationship between AIA volatility and various uncertainties is contingent upon the prevailing market volatility regime. Specifically, our findings indicate that AIA volatility exhibits sensitivity to uncertainties related to government policies and market risk measures only under a low-volatility regime. The overall results underscore the value of MSR and QR methodologies, as they uncover disparities in AIA volatilities across shifts in economic regimes. The observation that economic uncertainty indicators, such as EPU and VIX, play a diminished role in high volatility regimes aligns with intuitive expectations. During major volatility events, the primary drivers are often unrelated to prevailing market sentiment or economic conditions. Consequently, sentiment or uncertainty indicators exhibit weak associations with abrupt increases in volatility. Conversely, shifts in sentiment and uncertainty during periods of low volatility (when markets are calm) are more likely to evoke substantial investor responses.

Volatility holds a pivotal role in the pricing mechanisms of options and derivatives, influencing the dynamics of futures contract prices through its impact on the underlying asset's returns, leading to contango and backwardation. Our paper yields insights that are invaluable for investors, speculators, and other participants in equity markets, shedding light on the behavior and volatility patterns of AIA. This information equips market participants to formulate strategic hedging approaches, employing options and futures contracts for defensive positioning. Our results emphasize the informative nature of movements of uncertainty indices for crafting effective hedging strategies when investing in Ishares Asia 50 ETF.

Our contributions to the literature are twofold. Firstly, we affirm the significance of sentiment and uncertainties in shaping long-term volatility, advocating for their inclusion in statistical models that aim to elucidate volatility patterns, including phenomena like volatility clustering/persistence. Secondly, we employ a robust GARCH-MIDAS framework, complemented by Markov switching and quantile regressions, underscoring the framework's efficacy in capturing the regime-related dynamics of AIA volatilities. This two-step multivariate methodology deviates from most previous research that includes only two covariates at a time in the GARCH-MIDAS estimation, thus leading to potential specification errors.

Declarations

Authors declare no funding or conflict of interest

Data for the research are available upon request

**Acknowledgment**:

We are grateful to anonymous reviewers for their valuable comments and suggestions. Remaining errors are the authors' responsibility.